# Tachyonic Dirac Equation Revisited

Luca Nanni

Faculty of Natural Science, University of Ferrara, 44122 Ferrara, Italy

luca.nanni@edu.unife.it

**Abstract** In this paper, we revisit the two theoretical approaches for the formulation of the tachyonic Dirac equation. The first approach works within the theory of restricted relativity, starting from a Lorentz invariant Lagrangian consistent with a spacelike four-momentum. The second approach uses the theory of relativity extended to superluminal motions and works directly on the ordinary Dirac equation through superluminal Lorentz transformations. The equations resulting from the two approaches show mostly different, if not opposite, properties. In particular, the first equation violates the invariance under the action of the parity and charge conjugation operations. Although it is a good mathematical tool to describe the dynamics of a space-like particle, it also shows that the mean particle velocity is subluminal. In contrast, the second equation is invariant under the action of parity and charge conjugation symmetries, but the particle it describes is consistent with the classical dynamics of a tachyon. This study shows that it is not possible with the currently available theories to formulate a covariant equation that coherently describes the neutrino in the framework of the physics of tachyons, and depending on the experiment, one equation rather than the other should be used.

## 1. Introduction

The Dirac equation is one of the most widely used mathematical tools in modern theoretical physics [1-4]. It was formulated in 1928 and has since been mentioned in the literature of most scientific areas. In physics and quantum chemistry, the Dirac equation and the Schrodinger equation together form the main foundation. In its non-linear version [5-7], the Dirac equation is used in condensed matter physics [8-10] and quantum optics [11-14]. The Dirac equation has also been generalised for curved space-time in order to study the behaviour of fermions in gravitational fields [15-17]. In the framework of mathematical physics, the Dirac equation has been studied for different potentials (e.g., Coulomb, Yukawa, and Poschl-Teller) in order to test or find new methods of solving the associated Cauchy problem [18-20]. However, the Dirac equation has not been adequately considered in the framework of quantum physics of tachyonic fermions. This is mainly because its original formulation is not suitable to describe the dynamics of a superluminal particle that violates the Lorentz invariance. However, experiments performed on neutrinos show that this particle, under certain conditions, fulfils the energy-momentum relationship typical of tachyons [21]. In particular, in tritium beta decay measurements, the electron neutrino shows a negative mass-square ($m_{\nu_e} = -2.5 \pm 3.3\ eV^2$) [22]. A similar result was also obtained for the muon neutrino in pion decay

[23]. SN1987 data also has evidence for the superluminal behaviour of neutrinos [24]. In this context, there is a need to have a suitable covariant equation to describe the dynamics of ½-spin fermions, which would represent the first step in constructing a theory capable of interpreting the mentioned experimental results, especially for the purpose of designing new experiments that can provide more precise measurements of the neutrino mass-square.

In this study, two physical-mathematical approaches are adopted and reworked in order to formulate a Dirac-type equation consistent with the tachyonic energy-momentum relationship. By tachyonic Dirac-type equation we mean a covariant equation that describes, in the quantum mechanics framework, the dynamics of a ½-spin particle moving with superluminal velocity. The first approach is based on the Tanaka Lagrangian, which although Lorentz invariant is nonetheless consistent with the space-like energy-momentum relationship owing to the presence of the fifth current operator in the mass energy term. This approach was introduced in the 1980s by Chodos [25] and investigated recently by Jentschura [26]. The purpose of revisiting this approach in this study is to highlight the physical meaning of the results obtained by solving exactly the obtained tachyonic equation. In particular, this study proves that this equation describes a new type of particle that fulfils the tachyonic energy-momentum relationship but propagates at subluminal velocity. This unexpected result is related to the symmetry of the obtained wave functions and to the duality between tachyonic and ordinary spacetime. The second approach is instead based on the introduction of superluminal Lorentz transformations (SLTs) that are typical of the theory of relativity extended to superluminal motions (ETRR) [27]. The results obtained from the second approach are substantially different than those obtained from the first approach, showing that the resulting tachyonic equation in the second approach is more suitable for describing the dynamics of a particle that is closest to the classical concept of the tachyon. The behaviours of the solutions of the two obtained equations under the action of discrete symmetries (parity, time-reversal and charge-conjugation) are also investigated. In this aspect too, the two equations have substantial differences. The first equation describes the symmetries of the phenomena in which the hypothetical space-like neutrino is involved, whereas the second equation is the only choice if we accept that the neutrino propagates at superluminal velocity.

To our knowledge, these results have not been published previously. These findings could help formulate a suitable theory to interpret the experimental data mentioned above, even though current theoretical models fail to consistently reconcile the physics of tachyons with quantum mechanics.

## 2. Tachyonic Dirac Equation from the Tanaka Lagrangian

The Tanaka Lagrangian [28] formulated to describe the tachyonic neutrino is as follows (in natural units):

$$\mathcal{L}_t = i\bar{\psi}\gamma^5\gamma^\mu\partial_\mu\psi - m\bar{\psi}\psi, \qquad (1)$$

where $\bar{\psi} = \psi^\dagger\gamma^0$, $\gamma^5 = i\gamma^0\gamma^1\gamma^2\gamma^3$, $(\gamma^5)^2 = \mathbb{1}$, $\gamma^\mu$ are the ordinary gamma Dirac matrices and $m$ is the real part of the tachyon's imaginary rest mass $im$. $\bar{\psi}$ and $\psi$ are independent coordinates of the Lagrangian operator. As can be seen, the mass energy term of this Lagrangian is identical to that of Dirac, whereas the kinetic energy component changes because of the introduction of the $\gamma^5$ matrix (fifth current) which does not commute with the other gamma matrices. This feature ensures that the associated tachyon equation, obtained by taking variations with respect to the $\bar{\psi}$ coordinate, fulfils the tachyonic energy-momentum relationship. It should also be stressed that Lagrangian (1) is Lorentz invariant and therefore describes the dynamics of the superluminal neutrino without the need to extend the theory of relativity (TRR) to superluminal motions, thus avoiding the problem of having to introduce a privileged reference frame. This also allows us to write the tachyonic equation using the ordinary Dirac matrices that do not depend on the subluminal reference frame. The equation obtained from Lagrangian (1) is the tachyonic Dirac equation for a free particle and reads:

$$(i\hbar\gamma^0\partial_t - i\hbar c\gamma^k\partial_k - \gamma^5 mc^2)\psi = 0 \qquad (2)$$

From now on we set $k = 1,2,3$ and the zero index of the gamma matrix in the temporal term will be always written explicitly. Multiplying Equation (2) with its transpose conjugate gives the tachyonic Klein-Gordon equation:

$$(\hbar^2\partial_t^2 - \hbar^2 c^2 \nabla^2 + (im)^2 c^4)\psi = 0 \qquad (3)$$

Let us solve Equation (2) considering that, as is true for the ordinary Dirac equation, the wave functions are the product of four-spinors and a plane wave. For particle states, these functions are:

$$\psi^+(\uparrow) = \begin{pmatrix} 1 \\ 0 \\ u_1^+ \\ u_2^+ \end{pmatrix} e^{\left\{-\frac{i}{\hbar}(\boldsymbol{p}\cdot\boldsymbol{x} - Et)\right\}} \; ; \quad \psi^+(\downarrow) = \begin{pmatrix} 0 \\ 1 \\ u_3^+ \\ u_4^+ \end{pmatrix} e^{\left\{-\frac{i}{\hbar}(\boldsymbol{p}\cdot\boldsymbol{x} - Et)\right\}}, \qquad (4)$$

whereas for antiparticles, they are:

$$\psi^-(\uparrow) = \begin{pmatrix} u_1^- \\ u_2^- \\ 1 \\ 0 \end{pmatrix} e^{\left\{\frac{i}{\hbar}(\boldsymbol{p}\cdot\boldsymbol{x} - Et)\right\}} \; ; \quad \psi^-(\downarrow) = \begin{pmatrix} u_3^- \\ u_4^- \\ 0 \\ 1 \end{pmatrix} e^{\left\{\frac{i}{\hbar}(\boldsymbol{p}\cdot\boldsymbol{x} - Et)\right\}}, \qquad (5)$$

where $\boldsymbol{p}$ and $E$ are space-like momentum and energy respectively, given by:

$$E = \gamma imc^2 = |\gamma|mc^2 \; ; \quad p_j = \gamma imu_j = |\gamma|mc\beta_j \quad j = x, y, z, \qquad (6)$$

where $\gamma$ is the (imaginary) tachyonic Lorentz factor $\left[1 - (u_j/c)^2\right]^{-1/2}$ and $\beta$ is the relativistic coefficient $(u_j/c)$. Substituting the explicit form of the gamma matrices in Equation (2), we obtain a set of four linear equations:

$$\begin{pmatrix} i\hbar\partial_t & 0 & -i\hbar c\partial_z & -(i\hbar c\partial_x + \hbar c\partial_y + mc^2) \\ 0 & i\hbar\partial_t & (-i\hbar c\partial_x + \hbar c\partial_y - mc^2) & i\hbar c\partial_z \\ i\hbar c\partial_z & (i\hbar c\partial_x + \hbar c\partial_y - mc^2) & -i\hbar\partial_t & 0 \\ (i\hbar c\partial_x - \hbar c\partial_y - mc^2) & -i\hbar c\partial_z & 0 & -i\hbar\partial_t \end{pmatrix} \times$$

$$\times \psi^{(\pm)}(\uparrow\downarrow) e^{\{\mp\frac{i}{\hbar}(\mathbf{p}\cdot\mathbf{x}-Et)\}} = 0, \tag{7}$$

where $\psi^{(\pm)}(\uparrow\downarrow)$ is the spinor part of the wave function. By calculating all the partial derivatives, we get a set of algebraical equations by which the coefficients $u_{1,3}^{\pm}$ and $u_{2,4}^{\pm}$ are explicitly obtained:

$$\begin{cases} u_1^+ = p_z c/E \ ; \ u_2^+ = [c(p_x + ip_y) - mc^2]/E \\ u_3^+ = [c(p_x - ip_y) - mc^2]/E \ ; \ u_4^+ = -p_z c/E \\ u_1^- = p_z c/E \ ; \ u_2^- = [c(p_x + ip_y) - mc^2]/E \\ u_3^- = [c(p_x - ip_y) - mc^2]/E \ ; \ u_4^- = -p_z c/E \end{cases} \tag{8}$$

As usual, the wave functions must be normalised by imposing the condition $\langle\psi|\psi\rangle = 1$, which also helps obtain the normalisation factor $\mathfrak{N} = E/[2(E^2 + m^2c^4 - |\gamma|\beta_x m^2 c^4)]^{1/2}$. For simplicity, the normalisation factor is omitted in the following equations, as it does not contribute to the discussion of the results. To investigate the properties of the wave functions (4) and (5), we write the ψ matrix whose columns are the spinors of the tachyonic Dirac equation (2) and compare it with the equation below denoted by $\tilde{\psi}$, which is obtained by the spinors of the ordinary Dirac equation [29]:

$$\psi = \begin{pmatrix} 1 & 0 & \frac{p_z c}{E} & \frac{[c(p_x - ip_y) - mc^2]}{E} \\ 0 & 1 & \frac{[c(p_x + ip_y) - mc^2]}{E} & -\frac{p_z c}{E} \\ \frac{p_z c}{E} & \frac{[c(p_x - ip_y) - mc^2]}{E} & 1 & 0 \\ \frac{[c(p_x + ip_y) - mc^2]}{E} & -\frac{p_z c}{E} & 0 & 1 \end{pmatrix} \tag{9}$$

$$\tilde{\psi} = \begin{pmatrix} 1 & 0 & \frac{\tilde{p}_z c}{\tilde{E}+mc^2} & \frac{c(\tilde{p}_x - i\tilde{p}_y)}{\tilde{E}+mc^2} \\ 0 & 1 & \frac{c(\tilde{p}_x + i\tilde{p}_y)}{\tilde{E}+mc^2} & -\frac{\tilde{p}_z c}{\tilde{E}+mc^2} \\ \frac{\tilde{p}_z c}{\tilde{E}+mc^2} & \frac{c(\tilde{p}_x - i\tilde{p}_y)}{\tilde{E}+mc^2} & 1 & 0 \\ \frac{c(\tilde{p}_x + i\tilde{p}_y)}{\tilde{E}+mc^2} & -\frac{\tilde{p}_z c}{\tilde{E}+mc^2} & 0 & 1 \end{pmatrix} \tag{10}$$

The tilde above the energy and momentum denotes time-like observables. We can immediately recognise that the main difference between the non-trivial components of the two matrices $\psi$ and $\tilde{\psi}$ is represented by the change of position of the mass energy. This difference is due to the matrix $\gamma^5$ in Equation (2), which moves the mass terms on the secondary diagonal of the energy operator. This ensures that the Hermitian product Equation (2) times itself is consistent with the tachyonic energy-momentum relationship. The physical meaning of this difference lies in the duality between the time-like and space-like observables. In this regard, let us consider the tachyonic momentum from the classical framework $p = |\gamma|mu$. The superluminal velocity $u$ can be always written as $u = c^2/\tilde{u}$, where $\tilde{u} < c$. In other words, it is always possible to find a subluminal velocity $\tilde{u}$ such that $u$ is tachyonic. Each superluminal reference frame is in biunivocal correspondence with its bradyonic dual. Under this assumption, it is easy to verify that the momentum of a tachyon is directly related to the energy of the corresponding dual bradyon and vice versa [30]:

$$\begin{cases} pc = i(1 - (u/c)^2)^{-1/2}muc = (1 - (\tilde{u}/c)^2)^{-1/2}mc^2 = \tilde{E} \\ E = i(1 - (u/c)^2)^{-1/2}mc^2 = (1 - (\tilde{u}/c)^2)^{-1/2}m\tilde{u}c = \tilde{p}c \end{cases} \quad (11)$$

From Equations (11), one clearly sees that the kinetic component of the total energy of the particle in the subluminal reference frame is equal to the kinetic component in the tachyonic reference frame deprived of the mass energy, namely $\tilde{p}^2c^2 = p^2c^2 - m^2c^4$. This is the reason why in the non-trivial components of Matrix (9) only the total energy of the particle appears in the denominator and a negative mass energy is added to the numerator. This is the physical interpretation of how the fifth current operator works in Equation (2). Equations (11) are useful to find the physical correlation between the subluminal rest frame, where $\tilde{u} = 0$, and the superluminal transcend *rest* frame where $u \to \infty$:

$$\tilde{u} = 0 \Rightarrow u = \frac{c^2}{\tilde{u}} \to \infty \; ; \; \tilde{E} = mc^2 \Rightarrow p = \frac{\tilde{E}}{c} = mc \; ; \; \tilde{p} = 0 \Rightarrow E = \tilde{p}c = 0 \quad (12)$$

Matrices (9) and (10) converge when $u, \tilde{u} \to c$ and when the energy mass $mc^2 \to 0$. These limits represent the analytic continuity between the tachyonic and ordinary wave functions. This property is a consequence of the fact that the Lagrangian from which the ordinary and tachyon Dirac equations are obtained are both Lorentz invariant. Finally, one notes that the structures of the two matrices $\psi$ and $\tilde{\psi}$ are conserved. That is, there is a correspondence between a given subluminal state and a tachyonic one with the same spin projection and the same energy sign.

Let us now write the Hamiltonian operator associated with Equation (2):

$$H = i\hbar c \alpha^k \partial_k + \gamma^0 \gamma^5 mc^2 \quad (13)$$

This operator is not Hermitian because $\gamma^0\gamma^5$ is antihermitian. In the quantum mechanics of ordinary particles, all operators representing observables are Hermitian. In the quantum mechanics of tachyons, non-Hermitian operators work properly [26]. In fact, the eigenvalues (continuous spectrum) of the energy operator (13) are all real $E = \pm\sqrt{p^2c^2 - m^2c^4}$ and the time evolution operator is unitary. The necessary and sufficient condition is that the non-Hermitian Hamiltonian is $\mathcal{PT}$ invariant (spacetime inversion). Otherwise it would have imaginary eigenvalues, or the time-evolution operator would be non-unitary (with consequent non-conservation of probability) [31]. Let us clarify this aspect further. To better understand the algebraical structure of Matrix (13), let us break it down into the sum of four matrices:

$$H = i\hbar c \partial_x \gamma^5 \gamma^1 + i\hbar c \partial_y \gamma^5 \gamma^2 + i\hbar c \partial_z \gamma^5 \gamma^3 + mc^2 \mathbb{1} \quad (14)$$

The first three matrices at the right-hand side of Equation (14) are antihermitian whereas the latter is trivially Hermitian. However, the fifth current $\gamma^5$ does not commute with the other gamma matrices and this is the main difference between the ordinary and tachyonic equations. We define the self-adjoint operator as follows:

$$\mathcal{O}^\dagger = (\gamma^5 \mathcal{O} \gamma^5)^\dagger \tag{15}$$

The Hermicity of the tachyonic energy operator $H$ is recovered. The ordinary Dirac formalism is modified—this modification is inevitable to formulate a tachyonic equation without using superluminal transformations.

Let us now investigate the invariance of the tachyonic Dirac equation under the action of the parity $\mathcal{P}$, time-reversal $\mathcal{T}$ and charge-conjugation $\mathfrak{C}$ symmetry operations. These discrete symmetries work as follows: $\mathcal{P}\psi(\boldsymbol{x}, t) = \Lambda_\mathcal{P} \psi(-\boldsymbol{x}, t)$, $\mathcal{T}\psi(\boldsymbol{x}, t) = \Lambda_\mathcal{T} \gamma^0 \psi^*(\boldsymbol{x}, -t)$ and $\mathfrak{C}\psi(\boldsymbol{x}, t) = \Lambda_\mathfrak{C} \gamma^0 \psi^*(\boldsymbol{x}, t)$. Parity performs a reflection of the space coordinates at the origin, time-reversal operates on time coordinates and charge-conjugation transforms a particle in its antiparticle, and vice versa. Therefore, parity changes sign to position and momentum, and leaves spin unchanged; time-reversal changes sign to spin, momentum and velocity; and charge-conjugation changes sign to all *intrinsic* charges leaving unchanged spin, momentum and position. Since we proved the analytic continuity of the tachyonic wavefunctions, we expect that, as is true for the ordinary Dirac equation, the superluminal one too is invariant under these transformations. This can be illustrated by rewriting Equation (2) as:

$$(i\hbar \mathbb{1} \partial_t - i\hbar c \alpha^k \partial_k - \gamma^0 \gamma^5 mc^2)\psi = 0 \tag{16}$$

and comparing it with the ordinary Dirac one:

$$(i\hbar \mathbb{1} \partial_t - i\hbar c \alpha^k \partial_k - \gamma^0 mc^2)\tilde{\psi} = 0 \tag{17}$$

It is well-known that for the ordinary Dirac equation, parity symmetry is ensured by the operator $\Lambda_\mathcal{P} = \gamma^0$ [32]. In fact, this unitary operator transforms the space-inverted Dirac operator in itself:

$$(\gamma^0)^{-1}(i\hbar \gamma^0 \partial_t + i\hbar c \gamma^k \partial_k - \mathbb{1} mc^2)\gamma^0 = (i\hbar \gamma^0 \partial_t - i\hbar c \gamma^k \partial_k - \mathbb{1} mc^2) \tag{18}$$

Applying this transformation to Equation (2) we see that $\gamma^5$ changes sign, and parity is violated:

$$\begin{aligned}(\gamma^0)^{-1}(i\hbar \gamma^0 \partial_t + i\hbar c \gamma^k \partial_k - \gamma^5 mc^2)\gamma^0 = \\ = (i\hbar \gamma^0 \partial_t - i\hbar c \gamma^k \partial_k + \gamma^5 mc^2)\end{aligned} \tag{19}$$

However, comparing Equations (16) and (17), we see that the tachyonic mass operator, analogous to the Dirac one, is $\gamma^0 \gamma^5$. Using this matrix as the $\Lambda_\mathcal{P}$ operator and changing the sign in the right hand-side of Equation (19), the similarity transformation returns the tachyonic equation, since $(\gamma^0 \gamma^5)^{-1} \gamma^k (\gamma^0 \gamma^5) = -\gamma^k$ and $(\gamma^0 \gamma^5)^{-1} \gamma^5 (\gamma^0 \gamma^5) = \gamma^5$. Arbitrarily changing the sign of one side of the equality is always licit, as the similarity transformation introduces a phase factor that does not affect the result. The parity is then recovered but not in the usual Dirac framework. The same approach can be followed to study the time-reversal symmetry. For the ordinary Dirac equation, this symmetry is assured by the operator $\Lambda_{\mathcal{T}} = i\gamma^1\gamma^3$ [32]. Since the first terms of Equations (16) and (17) are identical, $\mathcal{T}$ symmetry is expected to be verified also by the tachyonic equation through the same operator $\Lambda_{\mathcal{T}}$. By performing the similarity transformation on the conjugate time-reversed tachyonic equation, one gets:

$$(i\gamma^1\gamma^3)^{-1}(i\hbar(\gamma^0)^*\partial_t + i\hbar c(\gamma^k)^*\partial_k - (\gamma^5)^*mc^2)i\gamma^1\gamma^3 = \\ = (i\hbar\gamma^0\partial_t - i\hbar c\gamma^k\partial_k - \gamma^5 mc^2) \quad (20)$$

In this case, similarity transformation reverses the sign of the $(\gamma^k)^*$ matrices leaving unchanged the signs of the other ones. This proves the time-reversal invariance of the tachyonic equation using the same $\Lambda_{\mathcal{T}}$ operator used in the ordinary Dirac theory. Finally, we study the charge-conjugation symmetry. For this purpose, Equation (2) is formulated including an external field. For an electrically charged tachyon in an electromagnetic field, the tachyonic equation can be obtained by minimal coupling:

$$\left(i\hbar\gamma^0\partial_t - c\gamma^k\left(i\hbar\partial_k - \frac{e}{c}A_k\right) - \gamma^5 mc^2 - \mathbb{1}e\varphi\right)\psi = 0, \quad (21)$$

where $(\varphi, A_k)$ is the four-potential. Equation (21) differs from the Dirac one only for the fifth current in the mass term. The symmetry $\mathfrak{C}$ holds if we find a transformation such that:

$$(\Lambda_{\mathfrak{C}})^{-1}\left(i\hbar(\gamma^0)^*\partial_t - c(\gamma^k)^*\left(i\hbar\partial_k + \frac{e}{c}A_k\right) + (\gamma^5)^*mc^2 - \mathbb{1}e\varphi\right)\Lambda_{\mathfrak{C}} = \\ = \left(i\hbar\gamma^0\partial_t - c\gamma^k\left(i\hbar\partial_k + \frac{e}{c}A_k\right) - \gamma^5 mc^2 - \mathbb{1}e\varphi\right) \quad (22)$$

For the ordinary Dirac equation in an external electromagnetic field, $\Lambda_{\mathfrak{C}}$ is given by $i\gamma^2$ (since only the $\gamma^2$ matrix changes sign under conjugation). However, this transformation does not change the sign of $\gamma^5$ and Equation (22) does not hold. Charge-conjugation is then violated, at least in the Dirac framework. In fact, if we use the new transformation $\Lambda_{\mathfrak{C}} = i\gamma^2\gamma^5$ the $\mathfrak{C}$ symmetry is recovered, provided that the sign of the right hand-side of Equation (22) gets reversed. Before going further, we check if the equation is also $\mathcal{PT}$ invariant, because, as we anticipated earlier, this symmetry is the condition that makes a non-Hermitian Hamiltonian an operator with real eigenvalues. Such a symmetry represents the inversion of spacetime coordinates. Therefore, we need find an operator $\Lambda_{\mathcal{PT}}$ that ensures the similarity transformation $(\Lambda_{\mathcal{PT}})^{-1}(-i\hbar\gamma^0\partial_t + i\hbar c\gamma^k\partial_k - \gamma^5 mc^2)\Lambda_{\mathcal{PT}} = (i\hbar\gamma^0\partial_t - i\hbar c\gamma^k\partial_k - \gamma^5 mc^2)$. For the ordinary Dirac equation, this operator is given by $\Lambda_{\mathcal{PT}} = i\gamma^5$. It is easy to prove that this operator works properly also on the tachyonic equation. Then, for the theorem proved in reference [31], the tachyonic non-Hermitian Hamiltonian has real eigenvalues and also ensures the unitary of the time-evolution operator.

In this section, we have studied Equation (2) supposing that it is endowed with superluminal velocity and satisfies the energy-momentum relationship $E^2 = p^2c^2 - m^2c^4$. This is proved for the energy-momentum relationship but must be verified for the velocity. In quantum mechanics, the velocity operator does not commute with the Hamiltonian and therefore we can calculate only its mean value. In the present study the particle velocity is computed as the group velocity of the tachyonic wave packet associated to equation (2), using the method proposed by Park [33]. This approach is based on the formulation of the equation of motion of the envelope function that characterizes the Gaussian wave packet. In order to simplify the calculation, we suppose the superluminal particle propagating along z-axis. Therefore, Equation (2) becomes:

$$[i\hbar\gamma^0\partial_t - i\hbar c\gamma^3\partial_z - mc^2\gamma^5]\psi = 0 \tag{23}$$

In this case the solutions are plane waves with positive (+) and negative (−) frequencies:

$$\begin{cases}\psi^+ = \Re\begin{pmatrix}u_1^+\\u_2^+\end{pmatrix}exp\{i(kz-\omega^+ t)\}\\ \psi^- = \Re\begin{pmatrix}u_1^-\\u_2^-\end{pmatrix}exp\{i(kz-\omega^- t)\}\end{cases} \tag{24}$$

where $u_i^\pm$ are the spinor components to find, $\Re$ is the normalization factor, while $k = p_z/\hbar$ and $\omega^\pm = E^\pm/\hbar$. Introducing the plan waves (24) in Equation (23) we get a set of algebraical equations through which the spinor components $u_i^\pm$ are obtained:

$$\begin{cases}u_1^+ = -E/(p_z c + mc^2) \; ; \; u_2^+ = E/(p_z c - mc^2)\\ u_1^- = E/(p_z c - mc^2) \; ; \; u_2^- = -E/(p_z c + mc^2)\end{cases} \tag{25}$$

Considering that $E = \gamma_t imc^2|\gamma_t|mc^2$ and $p_z = \gamma_t imu_z = |\gamma_t|mu_z$, where $\gamma_t = (1 - u^2/c^2)^{-1/2}$ is the pure imaginary tachyonic Lorentz factor and $u_z > c$, equations (25) can be rewritten as:

$$\begin{cases}u_1^+ = -|\gamma_t|c/(|\gamma_t|u_z + c) \; ; \; u_2^+ = |\gamma_t|c/(|\gamma_t|u_z - c)\\ u_1^- = |\gamma_t|c/(|\gamma_t|u_z - c) \; ; \; u_2^- = -|\gamma_t|c/(|\gamma_t|u_z + c)\end{cases} \tag{26}$$

while the normalization factor is $\Re = \sqrt{2(|\gamma_t|^2+1)}/(|\gamma_t|^2-1)$. We can now construct a wave packet by multiplying the obtained plane waves times a Gaussian envelope function $f^\pm(t,z)$ [33]:

$$\psi_G^\pm = \frac{\sqrt{2(|\gamma_t|^2+1)}}{(|\gamma_t|^2-1)}f^\pm(t,z)\begin{pmatrix}\frac{|\gamma_t|c}{(|\gamma_t|u_z\pm c)}\\ \mp\frac{\gamma_t c}{(|\gamma_t|u_z\mp c)}\end{pmatrix}exp\{i(k_0 z - \omega_0^\pm t)\} \tag{27}$$

Introducing the spinor (27) in the Equation (23) we get the two following differential equations for the envelope functions:

$$\begin{cases}\left(\frac{\partial}{\partial t} - c\frac{u_1^+}{u_1^-}\bigg|_{\gamma_0}\frac{\partial}{\partial z} + \frac{\Lambda_0^+}{i\hbar u_1^-}\bigg|_{\gamma_0}\right)f^+(t,z) = 0\\ \left(\frac{\partial}{\partial t} + c\frac{u_2^-}{u_2^+}\bigg|_{\gamma_0}\frac{\partial}{\partial z} + \frac{\Lambda_0^-}{i\hbar u_2^+}\bigg|_{\gamma_0}\right)f^-(t,z) = 0\end{cases} \tag{28}$$

where $\gamma_0$ is the module of the tachyonic Lorentz factor corresponding to the velocity $u_0$ ($u_0$ is the particle velocity at $t = 0$ and corresponds to the center of the Gaussian function), and $\Lambda_0^\pm = 2mc^2 u_1^\pm\big|_{\gamma_0}$. The numerical coefficient of the second term in Equations (28) is the group velocity $v_g$, which coincides with the propagation velocity of the wave packet [33]. Using Eq. (26) we obtain:

$$v_g = c \left.\frac{u_1^+}{u_1^-}\right|_{\gamma_0} = -c \left.\frac{u_2^-}{u_2^+}\right|_{\gamma_0} = \left(\frac{|\gamma_0|u_z - c}{|\gamma_0|u_z + c}\right) c \leq c \quad \forall u_o > c \qquad (29)$$

This quantity is always lower than the speed of light, which is an unexpected result. The tachyonic Dirac equation obtained from the Tanaka Lagrangian de facto describes a new type of subluminal particle, which can find its place within the Standard Model if a new symmetry is conjectured, a symmetry that transforms an ordinary particle (antiparticle) state into a corresponding one consistent with the tachyonic energy-momentum relationship [34]. This is precisely the starting hypothesis of the second approach that we will investigate next.

## 3. Tachyonic Dirac Equation through SLTs of Dirac Wavefunctions

In this section, the tachyonic Dirac equation is formulated by performing an SLT on the $\tilde{\psi}$ matrix. This approach implies that the tachyonic wavefunctions cannot be used to realise an irreducible representation of the Lorentz group. The SLT, denoted by $\Lambda_{STL}$, transforms a subluminal reference frame $\tilde{K}$ in a superluminal frame $K$ such that $(u - \tilde{u}) > c$. This operator must satisfy the following requirements:

$$(\Lambda_{STL})^\dagger = (\Lambda_{STL})^{-1} \quad and \quad \gamma^0 \Lambda_{STL} \gamma^0 = \Lambda_{STL}, \qquad (30)$$

so that the tachyonic spinor matrix can be written as $\psi = \Lambda_{STL}\tilde{\psi}$. The working of SLT is detailed in reference [27]. For our purpose, we employ the explicit form of $\Lambda_{STL}$ that was proposed by Lemke (but can be obtained from any other theory concerning the generalised theory of relativity [27]). It reads [35]:

$$\Lambda_{STL} = (\mathbb{1} + i\boldsymbol{\alpha} \cdot \boldsymbol{n})/\sqrt{2}, \qquad (31)$$

where $\boldsymbol{\alpha}$ is the vector whose components are the three Dirac matrices $\alpha^k$, and $\boldsymbol{n}$ is the vector whose components are the direction cosines that determine the relative direction of motion of the reference frame $\tilde{K}$ with respect to $K$. This matrix is unitary and Hermitian since $\mathbb{1}$ is symmetric and $\boldsymbol{\alpha} \cdot \boldsymbol{n}$ is antisymmetric. For simplicity, suppose that the relative motion between the two reference frames takes place along the $x$ axis. In this case, the $\Lambda_{STL}$ matrix is $\Lambda_{STL} = [\mathbb{1} + i\gamma^5]/\sqrt{2}$. Applying this transformation on Matrix (10), we get:

$$\psi = \frac{1}{\sqrt{2}} \begin{pmatrix} 1 + i\frac{\tilde{p}_x c}{\tilde{E} + mc^2} & 0 & 0 & i + \frac{\tilde{p}_x c}{\tilde{E} + mc^2} \\ 0 & 1 + i\frac{\tilde{p}_x c}{\tilde{E} + mc^2} & i + \frac{\tilde{p}_x c}{\tilde{E} + mc^2} & 0 \\ 0 & i + \frac{\tilde{p}_x c}{\tilde{E} + mc^2} & 1 + i\frac{\tilde{p}_x c}{\tilde{E} + mc^2} & 0 \\ i + \frac{\tilde{p}_x c}{\tilde{E} + mc^2} & 0 & 0 & 1 + i\frac{\tilde{p}_x c}{\tilde{E} + mc^2} \end{pmatrix} \qquad (32)$$

From Matrix (32), we note that the structure of transformed spinors is different than the subluminal one (see Matrix (10), with the $y$ and $z$ component set to zero). However, for the limit $\tilde{u} \to c$ (or $mc^2 \to 0$), the transformed (tachyonic) wave functions converge to the ordinary ones, less than a phase factor of $(1 \pm i)/\sqrt{2}$. We can therefore say that analytic continuity is fulfilled even in this approach. However, whereas in the previous case this property is a direct consequence of the fact that the tachyonic equation is derived from a Lorentz invariant Lagrangian, in this case the analytic continuity is due to the fact that the equation is obtained by superluminal transformation of the ordinary Dirac one.

To obtain the tachyonic equation, we perform the following transformation on the Dirac spinor:

$$\psi = \Lambda_{STL}\tilde{\psi} \quad \Rightarrow \quad \tilde{\psi} = (\Lambda_{STL})^{-1}\psi = \frac{1}{\sqrt{2}}(\mathbb{1} - i\boldsymbol{\alpha}\cdot\boldsymbol{n})\psi \tag{33}$$

Since $\tilde{\psi}$ is a solution of the ordinary Dirac equation, we can write:

$$(i\hbar\gamma^0\partial_t - i\hbar c\gamma^k n_k \partial_k - \mathbb{1}mc^2)\frac{1}{\sqrt{2}}(\mathbb{1} - i\boldsymbol{\alpha}\cdot\boldsymbol{n})\psi = 0, \tag{34}$$

where for consistency of formalism we used the scalar product between the $\boldsymbol{\alpha}$ and $\boldsymbol{n}$ vectors. Multiplying the left-hand side with $(\Lambda_{STL})$, we get:

$$\frac{1}{2}(\mathbb{1} + i\boldsymbol{\alpha}\cdot\boldsymbol{n})(i\hbar\gamma^0\partial_t - i\hbar c\gamma^k n_k \partial_k - \mathbb{1}mc^2)(\mathbb{1} - i\boldsymbol{\alpha}\cdot\boldsymbol{n})\psi = 0 \tag{35}$$

Performing the matrix products, we see that:

$$\Lambda_{STL}\gamma^0(\Lambda_{STL})^{-1} = i\gamma^k n_k \quad ; \quad \Lambda_{STL}\gamma^k n_k(\Lambda_{STL})^{-1} = i\gamma^0 \tag{36}$$

As expected, the superluminal transformation interchanges the temporal variable with the spatial one and vice versa, that is, operator $\partial_t$ is interchanged with $\partial_k$. Therefore, substituting relations (36) in Equation (35), we arrive at the tachyonic Dirac equation:

$$(\hbar\gamma^0\partial_t - \hbar c\gamma^k n_k \partial_k + \mathbb{1}mc^2)\psi = 0 \tag{37}$$

Equation (37) is the superluminal transformation of the Dirac one. Its covariance is trivially inherent from the fact that superluminal transformations are the components of the Lorentz group ETRR which, in turn, is consistent with the postulates of the TRR [35]. In other words, a superluminal transformation transforms in a covariant way any other equation that satisfies the formalism of the ordinary TRR.

Let us rewrite Equation (31) as:

$$(i\hbar\mathbb{1}\partial_t - i\hbar c\alpha^k n_k \partial_k + i\gamma^0 mc^2)\psi = 0 \tag{38}$$

The Hamiltonian associated with Equation (37) is:

$$H = i\hbar c\alpha^k n_k \partial_k - i\gamma^0 mc^2 \tag{39}$$

This Hamiltonian is not Hermitian since the operator $i\gamma^0$ is antihermitian but the eigenvalues are reals. The main difference between the Hamiltonians (39) and (13) is represented by the matrix $-i\gamma^0$ that replaces the antihermitian operator $\gamma^0\gamma^5$.

Let us now investigate the behaviour of Equation (37) under the action of discrete symmetries. For convenience, let us rewrite it as:

$$(i\hbar\gamma^0\partial_t - i\hbar c\gamma^k\partial_k + i\mathbb{1}mc^2)\psi = 0, \tag{40}$$

where the cosine directors have been omitted to simplify the formalism. Concerning parity, performing the similarity transformation on Equation (40) with inverted space coordinates by $\Lambda_\mathcal{P} = \gamma^0$, the operator $(i\hbar\gamma^0\partial_t - i\hbar c\gamma^k\partial_k + i\mathbb{1}mc^2)$ is obtained.

Therefore, Equation (40) is parity invariant, as is true for the ordinary Dirac equation. The action of time-reversal symmetry, however, presents some surprises. In fact, operating the similarity transformation using $\Lambda_{\mathcal{T}} = i\gamma^1\gamma^3$ on the conjugate time-reversed Equation (40) we get the operator $(i\hbar\gamma^0\partial_t - i\hbar c\gamma^k\partial_k - i\mathbb{1}mc^2)$. This is because the transformation $\Lambda_{\mathcal{T}}^{-1}(i\mathbb{1})\Lambda_{\mathcal{T}}$ does not change the sign of the mass energy operator. Therefore, time-reversal symmetry is violated, and it is not possible to find any combination of gamma Dirac matrices that are able to verify the equality between Equation (40) and its time-reversed conjugate. Finally, we verify the action of the charge-conjugation symmetry, rewriting Equation (22) as follows:

$$(\Lambda_{\mathfrak{C}})^{-1}\left(i\hbar(\gamma^0)^*\partial_t - c(\gamma^k)^*\left(i\hbar\partial_k + \frac{e}{c}A_k\right) + (i\mathbb{1})^*mc^2 - \mathbb{1}e\varphi\right)\Lambda_{\mathfrak{C}} = \\ = \left(i\hbar\gamma^0\partial_t - c\gamma^k\left(i\hbar\partial_k + \frac{e}{c}A_k\right) + i\mathbb{1}mc^2 - \mathbb{1}e\varphi\right) \quad (41)$$

This transformation holds using the same operator that leaves unchanged the ordinary Dirac equation, namely $\Lambda_{\mathfrak{C}} = i\gamma^2$. Equation (38) therefore fulfils the charge-conjugation invariance as the Dirac one does. Regarding the behavior of the equation under the $\mathcal{PT}$ symmetry operation, it is simply verified that it is ensured by the operator $\Lambda_{\mathcal{PT}} = i\gamma^5$, which leaves the matrix $(i\mathbb{1})$ unchanged but inverts all the others. This result was expected because Hamiltonian (39) is non-Hermitian but with real eigenvalues.

To complete the comparison with the theory investigated in the previous section, we calculate the particle velocity described by Equation (38) using always the approach based on the tachyonic wave packet that leads to Equations (28). Even in this case, to simplify the calculation we consider a one-dimensional motion along z-axis. Since the structure of Equation (40) is the same as that of Dirac for an ordinary particle, the spinor components of the positive and negative frequency plane waves become:

$$\begin{cases} u_1^+ = -(|\gamma_t|u_z + c)/|\gamma_t|c \ ; \ u_2^+ = (|\gamma_t|u_z - c)/|\gamma_t|c \\ u_1^- = (|\gamma_t|u_z - c)/|\gamma_t|c \ ; \ u_2^- = -(|\gamma_t|u_z + c)/|\gamma_t|c \end{cases} \quad (42)$$

Substituting these components in Equation (29) we get:

$$v_g = c\frac{u_1^+}{u_1^-}\bigg|_{\gamma_0} = -c\frac{u_2^-}{u_2^+}\bigg|_{\gamma_0} = \left(\frac{|\gamma_0|u_z + c}{|\gamma_0|u_z - c}\right)c \geq c \quad \forall u_o > c \quad (43)$$

Hence, the equation obtained by superluminal transformation of the Dirac one describes a particle whose dynamics are consistent with the classical tachyon, namely faster than light velocity and tachyonic energy-momentum relationship being fulfilled. This is a consequence of the fact that the mass operator of Equation (38) does not change the structure of the ordinary Dirac Hamiltonian unlike what happens instead for the Hamiltonian obtained from the Tanaka Lagrangian.

**4. Discussion**

This paper shows that, depending on the approach used, tachyonic equations that are obtained describe different particles. The theory developed in Section 2 was formulated working within the TRR, whereas the second theory was derived within the framework of the ETRR. As deeply investigated by Sudarshan, Bilaniuk, Recami and others [36-38], the ETRR predicts the invariance of the speed of light and ensures that the laws of physics are the same in all inertial reference frames (subluminal and superluminal). Causal paradoxes may also be solved, at least in microphysics [39]. However, despite this, the equation obtained by performing a superluminal transformation of the ordinary Dirac equation leads to solutions with (almost) opposite properties to those characterising the tachyonic equation derived from the Lorentz invariant Lagrangian of Tanaka. In fact, the latter fulfils the tachyonic energy-momentum relationship because of the introduction of the fifth current operator in the mass energy term. The consequence is the loss of invariance under the action of parity and charge-conjugation symmetries, at least in the Dirac framework. Only the time-reversal invariance is preserved. This is because, as proved, Equation (2) describes the dynamics of a subluminal particle. In other words, we are dealing with a new type of ordinary particle, called by Salesi as the pseudo-tachyon [34], which arises from having forced an invariant Lorentz equation to fulfil the tachyonic energy-momentum relationship. The subluminal velocity of the pseudo-tachyon also explains why Equation (2) preserves the time-reversal invariance. However, it is surprising how the parity and charge-conjugation invariance can be recovered by modifying the ordinary operators representing the $\mathcal{P}$ and $\mathfrak{C}$ symmetries. In fact, the addition of the fifth current to these operators ensures the invariance of Equation (2), which explains why Equation (2) is widely used to describe a space-like neutrino even if its velocity still remains subluminal.

Things are different when the tachyonic equation is formulated by superluminal transformation of the ordinary Dirac equation. Equation (38) preserves many of the properties of the ordinary Dirac equation. For instance, the mean velocity of the particle is consistent with the space-like four-momentum (just like the ordinary Dirac equation is consistent with subluminal velocity and time-like four-momentum), and the tachyonic equation is invariant under parity and charge-conjugation symmetries. However, it remains to be explained why the time-reversal symmetry is violated without any possibility of recovering it by changing the form of the respective operators. We believe that the violation of this symmetry is due to the causality violation typical of superluminal motion. Even if the reinterpretation principle introduced by Feinberg, Sudarshan and others has solved this paradox [40], it is not possible to find an operator $\Lambda_{\mathfrak{C}}$ so as to ensure the time-reversal invariance of Equation (38).

We conclude by stating that currently, it is not possible to coherently formulate a covariant equation able of describing the hypothetical superluminal behaviour of the neutrino. If we want to remain in the orthodoxy of the TRR, then we have a good theory for describing a pseudo-tachyon neutrino. If this theory is accepted, then the experimental research should be directed towards the precision measurements of the square mass of the neutrino, rather than towards measuring the travel time between the source and the detector (because the particle is subluminal). On the other hand, if we enter into a more speculative framework, such as the ETRR, then we have an equation that does not adapt properly to the symmetries that characterise the phenomena in which the neutrino is involved. The use of one theory rather than the other depends therefore on the experiment under consideration. The enigmatic nature of the neutrino seems to inevitably affect the ability of theoretical physics to find a complete and satisfactory theory in the tachyonic framework.

**References**

[1] M. Berry, *Phys. World* **11**(2), 36 (1998).
[2] P.A.M. Dirac, *Proc. R. Soc. A* **117**, 610 (1928).
[3] B. Thaller, *The Dirac Equation*, Springer-Verlag Berlin Heidelberg (1992).
[4] F. Wilczek, *Int. J. Mod. Phys. A* **19**, 45 (2004).
[5] M.J. Esteban, E. Séré, *Discrete Cont. DYN-S.* **8**(2), 381 (2002).
[6] A.F. Ranada, *Classical Nonlinear Dirac Field Models of Extended Particles*, Quantum Theory Group, Fields and Particles, A.O. Barut ed., Reidel, Amsterdam (1982).
[7] A. Benhassine, *J. Math. Phys.* **60**(1), 011510 (2019).
[8] L.H. Haddad, L.D. Carr, *New J. Phys.* **17**, 093037 (2015).
[9] F. Dalfovo, S. Giorgini, L.P. Pitaevskii, S. Stringari, *Rev. Mod. Phys.* **71**, 463 (1999).
[10] P.G. Kevrekidis, D.J. Frantzeskakis, R. Carretero-Gonzalez, *Emergent Nonlinear Phenomena in Bose-Einstein Condensates*, Springer-Verlag, Berlin (2008).
[11] L.M. Kovachev, *Proceedings of Institute of Mathematics of NAS of Ukraine* **50**(2), 843 (2004).
[12] S.M. Barnett, *New J. Phys.* **16**, 093008 (2014).
[13] T.L. Silva, E.R.F. Taillebois, R.M. Gomes, S.P. Walborn, A.T. Avelar, *Phys. Rev. A* **99**, 022332 (2019).
[14] S.A.R. Horsley, *Phys. Rev. A* **98**, 043837 (2018).
[15] M.D. Pollock, *Acta Phys. Polon. B* **41**, 1827 (2010).
[16] M.D. de Oliveira, A.G.M. Schmidt, *Ann. Phys.* **401**, 21 (2019).
[17] M. Cariglia, *Classical Quant. Grav.* **21**(4), 1051 (2004).
[18] H. Tezuka, *AIP Adv.* **3**, 082135 (2013).
[19] M.R. Setare, S. Haidari, *Phys. Scr.* **81**(6), 065201 (2010).
[20] R.R. Hartmann, M.E. Portnoi, *Sci. Rep.* **7**, 11599 (2017).
[21] R. Ehrlich, *Astropart. Phys.* **66**, 11 (2015).
[22] D.E. Groom, Particle Data Group, Published in *Eur. Phys. J. C* **15**, 350 (2000).
[23] G. Cacciapaglia, A. Deandrea, L. Panizzi, *J. High En. Phys.* **2011**, 137 (2011).
[24] G. Pagliaroli, F. Vissani, M.L. Costantini, A. Ianni, *Astropart. Phys.* **31**, 163 (2009).
[25] A. Chodos, A. Hauser, V.A. Kostelecky, *Phys. Lett. B* **150**, 431 (1985).
[26] U.D. Jentschura, B.J. Wundt, *J. Phys. A-Math. Theor.* **45**, 444017 (2012).
[27] G.D. Maccarone, E. Recami, *Fund. Phys.* **14**, 367 (1984).
[28] S. Tanaka, *Prog. Theor. Phys.* **24**, 177 (1960).
[29] C. Amsler, *Nuclear and Particle Physics*, IOP Publishing Ltd, London (2015).
[30] A. Gusak, *Cherkasy University Bulletin: Physical and Mathematical Sciences* **16**(309) 111 (2014).
[31] C.M. Bender, *Rep. Prog. Phys.* **70**, 947 (2007).

[32] F. Schwabl, *Symmetries and Further Properties of the Dirac Equation. In: Advanced Quantum Mechanics*, Advanced Texts in Physics. Springer, Berlin, Heidelberg (2004).
[33] S.T. Park, *Phys. Rev. A* **86** 062105 (2012).
[34] G. Salesi, *Int. J. Mod. Phys. A* **28**, 5103 (1997).
[35] H. Lemke, *Il Nuovo Cimento* **35**(2), 181 (1976).
[36] E. Recami, *Rivista del Nuovo Cimento* **4**(2), 209 (1974).
[37] O.M. Bilaniuk, E.C.G. Sudarshan, *Nature* **223**, 386 (1969).
[38] S.K. Bose, *J. Phys.: Conf. Ser.* **196**, 012022 (2009).
[39] E. Recami, F. Fontana, R. Garavaglia, *Int. J. Mod. Phys. A* **15**(18), 2793 (2007).
[40] Y. Aharonov, A. Komar and L. Susskind, *Phys. Rev.* **182**, 1400 (1969).